# Bound state solutions of D-dimensional Schrödinger equation with Eckart potential plus modified deformed Hylleraas potential

Akpan N.Ikot[1], Oladunjoye A.Awoga[2] and Akaninyene D.Antia[3]

Theoretical Physics Group, Department of Physics, University of Uyo, Uyo, Nigeria.

[1] e-mail:ndemikot2005@yahoo.com, [2] e-mail:ola.awoga@yahoo.com, [3] e-mail:antiacauchy@yahoo.com

## Abstract

We study the D-dimensional Schrödinger equation for Eckart plus modified deformed Hylleraas potentials using the generalized parametric form of Nikiforov-Uvarov method. We obtain energy eigenvalues and the corresponding wave function expressed in terms of Jacobi polynomial. We also discussed two special cases of this potential comprises of the Hulthen potential and the Rosen-Morse potential in 3-Dimensions. Numerical results are also computed for the energy spectrum and the potentials,

## 1. Introduction

Exact or approximate solutions of Schrödinger equation has been solved in 3D for various potential [1-9,12-19]. Though few cases of these potentials are exactly solvable, for example the Coulomb and harmonic potentials [1,2]. However, various approximation methods have been developed by different authors to solve bound state solutions of the Schrodinger equation in recent times [3, 4]. This approximation is introduced for the centrifugal term in the Schrodinger equation [5- 7] for an arbitrary $l$-state. Recently, the studies of exponential-type potentials have attracted the attention of many researchers [40]. The potential under investigation includes the Hulthen potential [34],Maninng-Rosen potential[41] and the Eckart-like potential[42].

The purpose of this paper is to use a new approximation Scheme proposed in ref.[35] to the centrifugal term and apply the new scheme to study the Schrodinger equation for Eckart plus modified Hyllerras potentials in D-dimensions. The Eckart potential which has been studied by many researchers [8,9] is one of the most important exponential-type potential in physics and chemical physics whereas Hylleraas potential can be used to study diatomic molecules [10,11].

Recently, researchers have developed interest in the solutions of Schrödinger equation in D-dimensions. This is borne out of the desire to generalize the solution to multi-dimensional space. Such problems has been solved for both relativistic and non-relativistic cases with different potentials [12- 19].

The organization of the paper is organized as follows. In section2, we give a brief review of the Nikiforov-Uvarov method in its parametric form. Section 3 is devoted to Schrödinger equation in D-space. In section 4, we present the bound state solution of the Schrodinger equation in D-dimension. Finally, we give a brief conclusion in section 5.

## 2. Review of Nikiforov-Uvarov method and Its parametric form

The Nikiforov-Uvarov method (NU) based on the solution of a generalized second order linear differential equation with special orthogonal function [29, 30]. The Schrödinger equation

$$\psi''(r) + [E - V(r)]\psi(r) = 0 \qquad (1)$$

can be solved by the method by transforming this equation into by hypergeometric-type using the transformation, $s = s(x)$ to give

$$\psi''(s) + \frac{\bar{\tau}(s)}{\sigma(s)}\psi'(s) + \frac{\bar{\sigma}(s)}{\sigma^2(s)}\psi(s) = 0 \qquad (2)$$

In order to find the solution to (2) we set the wave function as

$$\psi(s) = \varphi(s)\chi_n(s) \qquad (3)$$

Substituting Eq. (3) into Eq. (2) reduces Eq. (2) into hypergeometric-type equation

$$\sigma(s)\chi_n''(s) + \tau(s)\chi_n'(s) + \lambda\chi_n(s) = 0 \qquad (4)$$

where the wave function $\varphi(s)$ is a logarithmic function

$$\frac{\varphi'(s)}{\varphi(s)} = \frac{\pi(s)}{\sigma(s)} \qquad (5)$$

and the other wave function $\chi(s)$ is the hypergeometric-type function on whose polynomials satisfies the Rodrigues relation

$$\chi_n(s) = \frac{B_n}{\rho(s)} \frac{d^n}{ds^n} [\sigma^n(s)\rho(s)] \qquad (6)$$

where $B_n$ is the normalization constant and the weight function $\rho(s)$ satisfies the condition

$$[\sigma(s)\rho(s)]' = \tau(s)\rho(s) \qquad (7)$$

The required $\pi(s)$ and $\lambda$ for the NU method are defined as

$$\pi(s) = \frac{\sigma' - \tilde{\tau}}{2} \pm \sqrt{\left(\frac{\sigma' - \tilde{\tau}}{2}\right)^2 - \tilde{\sigma}(s) + k\sigma(s)} \qquad (8)$$

and

$$\lambda = k + \pi'(s) \qquad (9)$$

respectively. It is necessary that the term under the square root sign in Eq. (8) be the square of a polynomial. To calculate $k$ from Eq. (8) then the discriminant of the quadratic term must vanish. The eigenvlues in Eq. (9) take the form

$$\lambda = \lambda_n = -n\tau'(s) - \frac{n(n-1)}{2}\sigma''(s), n = 0, 1, 2 \ldots \ldots \quad (10)$$

where

$$\tau(s) = \bar{\tau}(s) + 2\pi(s) \quad (11)$$

and its derivative is less than zero is the necessary condition for bound state solutions. The energy eigenvalues are obtained by comparing Eq. (9) and Eq. (10).

The parametric generalization of the NU method that is valid for both central and non-central exponential-type potential [31] can be derived by comparing the generalized hypergeometric-type equation.

$$\psi''(s) + \frac{c_1 - c_2 s}{s(1 - c_3 s)}\psi'(s) + \frac{1}{s^2(1 - c_3 s)^2}[-\xi_1 s^2 + \xi_2 s - \xi_3]\psi(s) = 0 \quad (12)$$

with Eq. (2) and we obtain the following parametric polynomials.

$$\bar{\tau}(s) = (c_1 - c_2 s) \quad (13)$$

$$\bar{\sigma}(s) = -\xi_1 s^2 + \xi_2 s - \xi_3 \quad (14)$$

$$\sigma(s) = s(1 - c_3 s) \quad (15)$$

Substituting Eq. (13) – Eq. (15) into Eq. (8), we get

$$\pi(s) = c_4 - c_5 s \pm \sqrt{[(c_6 - c_3 k_\pm)s^2 (c_7 + k_\pm)s + c_8]} \quad (16)$$

where

$$c_4 = \frac{1}{2}(1 - c_1), c_5 = \frac{1}{2}(c_2 - 2c_3), c_6 = c_5^2 + \xi_1$$

$$c_7 = 2c_4 c_5 - \xi_2, c_8 = c_4^2 + \xi_3 \quad (17)$$

We obtain the parametric $k_\pm$ from the condition that the function under the square root should be the square of a polynomial,

$$k_\pm = -(c_7 + 2c_3c_8) \pm 2\sqrt{c_8c_9} \tag{18}$$

where

$$c_9 = c_3c_7 + c_3^2c_8 + c_6 \tag{19}$$

Hence the function $\pi(s)$ in Eq. (16) becomes

$$\pi(s) = c_4 + c_5 s - [(\sqrt{c_9} + c_3\sqrt{c_8})s - \sqrt{c_8}] \tag{20}$$

The negative $k_-$ value is obtained as

$$k_- = -(c_7 + 2c_3c_8) - 2\sqrt{c_8c_9} \tag{21}$$

Thus, from the relation Eq. (11), we have

$$\tau(s) = c_1 + 2c_4 - (c_2 - 2c_5)s - 2[(\sqrt{c_9} + c_3\sqrt{c_8})s - \sqrt{c_8}] \tag{22}$$

whose derivative must be negative, that is

$$\tau'(s) = -2c_3 - 2[(\sqrt{c_9} + c_3\sqrt{c_8}) < 0] \tag{23}$$

solving Eq. (9) and Eq. (10), we obtain the parametric energy equation as

$$(c_2 - c_3)n + c_3n^2 - (2n + 1)c_5 + (2n + 1)[\sqrt{c_9} + c_3\sqrt{c_8}] + c_7$$
$$+2c_3c_8 + 2\sqrt{c_8c_9} = 0 \tag{24}$$

The weight function is obtain as

$$\rho(s) = s^{c_{10}}(1 - c_3s)^{c_{11}} \tag{25}$$

And together with Eq. (6), we obatin

$$\chi_n(s) = P_n^{(c_{10},c_{11})}(1 - 2c_3 s) \tag{26}$$

Where $P_n^{(c_{10},c_{11})}$ are the Jacobi polynomials. For special cases with $c_3 = 0$, the Jacobi polynomial reduces to Laggurre polynomials.

where

$$c_{10} = c_1 + 2c_4 + 2\sqrt{c_8} \tag{27}$$

$$c_{11} = 1 - c_1 - 2c_4 + \frac{2}{c_3}\sqrt{c_9} \tag{28}$$

The other part of the wave function is obtained from Eq. (5) as

$$\varphi(s) = s^{c_{12}}(1 - c_3 s)^{c_{13}} \tag{29}$$

where

$$c_{12} = c_4 + \sqrt{c_8} \tag{30}$$

$$c_{13} = -c_4 + \frac{1}{c_3}\left(\sqrt{c_9} - c_5\right) \tag{31}$$

Thus, the total wave function becomes

$$\psi_n(s) = N_n s^{c_{12}}(1 - c_3 s)^{c_{13}} P_n^{(c_{10}-c_{11})}(1 - 2c_3 s) \tag{32}$$

where $N_n$ is the normalization constant.

## 3. Schrödinger equation in D-dimensions

The SE presented in Eq. (1) is in 3-D. For D-dimensional space, the SE is [32].

$$-\frac{\hbar^2}{2\mu}[\nabla_D^2 + V(r)]\psi_{nlm}(r, \Omega_D) = E_{n,l}\psi_{nlm} \tag{33}$$

where the Laplacian operator is

$$\nabla_D^2 = \frac{1}{r^{D-1}} \frac{\partial}{\partial r}\left(r^{D-1} \frac{\partial}{\partial r}\right) - \frac{\Lambda_D^2}{r^2}(\Omega_D) \qquad (34)$$

The second term of Eq. (34) is the multidimensional space centrifugal term. $\Omega_D$ represents the angular coordinates. In this line, the operator $\Lambda_D^2$ yields hyperspherical harmonic as its eigenfunction. This helps us to write

$$\psi_{nlm}(r,\Omega_D) = R_{n,l}(r) Y_l^m(\Omega_D) \qquad (35)$$

$R_{n,l}$ is the radial part of the equation and $Y_l^m(\Omega_D)$ is the angular part called hyperspherical harmonics. The $Y_l^m(\Omega_D)$ obey the eigenvalue equation

$$\Lambda_D^2 Y_l^m(\Omega_D) = l(l+D-2) Y_l^m(\Omega_D) \qquad (36)$$

Substituting Eqs (34) and (35) into Eq. (33) and making use of Eq. (36) we obtain

$$\frac{1}{r^{D-1}} \frac{\partial}{\partial r}\left(r^{D-1} \frac{\partial R(r)}{\partial r}\right) + \frac{2\mu}{\hbar^2}\left[E_{n,l} - \frac{l(l+D-2)}{r^2} - V(r)\right] R(r) = 0 \qquad (37)$$

Eq. (37) is the SE in D-dimensions space.

## 4. Bound state solutions of Schrodinger equation

The potential Eckart plus Hylleraas is defined as [8,9,10,11]

$$V(r) = \frac{V_0}{b}\left(\frac{a - e^{-2\alpha r}}{1 - e^{-2\alpha r}}\right) - V_1 \frac{e^{-2\alpha r}}{1 - e^{-2\alpha r}} + V_2 \frac{e^{-2\alpha r}}{(1 - e^{-2\alpha r})^2} \qquad (38)$$

$V_0, V_1$ and $V_2$ are the depths of the potential well, $a$ and $b$ Hylleraas parameters and $\alpha$ is the inverse of the range of the potential. If $V_1 = V_2 = $

0, Eq. (38) reduces to deformed Hylleraas potential [10,11] and when $V_0 = 0 \; or \; b \to \infty$, Eq. (38) reduces to Eckart potential [8,9]. We display in Figs 1-3 the plots of the deformed Hylleraas potential, Eckart potential and the combined potentials $V(r)$ of Eq.(38) respectively.

The form of Eq. (37) is not suitable for bound state solutions, so we choose

$$R(r) = r^{-\left(\frac{D-1}{2}\right)} U(r) \tag{39}$$

Putting Eqs (39) and (38) into Eq. (37), we obtain

$$U''(r) + \left\{ \begin{array}{c} \frac{2\mu}{\hbar^2}\left[E - \frac{V_0}{b}\left(\frac{a - e^{-2\alpha r}}{1 - e^{-2\alpha r}}\right) + V_1 \frac{e^{-2\alpha r}}{1 - e^{-2\alpha r}} - V_2 \frac{e^{-2\alpha r}}{(1 - e^{-2\alpha r})^2}\right] \\ -\left(\frac{(D-1)(D-3)}{4} + l(l + D - 2)\right)\frac{1}{r^2} \end{array} \right\} U(r) = 0 \tag{40}$$

Eq. (40) can only be solved exactly in the s-wave for this potential. In order to find the approximate solutions of the radial Schrodinger equation of Eq. (40), we invoke use an approximation for the centrifugal [5, 6]

$$\frac{1}{r^2} = \frac{\alpha^2 e^{-2\alpha r}}{(1 - e^{-2\alpha r})^2} \tag{41}$$

This approximation is good for small values of the parameter $\alpha$ but fails for large values [33,34]. To accommodate large values of $\alpha$, we use a newly improves approximation scheme [35]

$$\frac{1}{r^2} = \frac{\omega e^{-2\alpha r}}{1 - e^{-2\alpha r}} + \frac{\lambda e^{-2\alpha r}}{(1 - e^{-2\alpha r})^2} \tag{42}$$

where $\omega$ and $\lambda$ are adjustable dimensionless parameters. In order to show that Eq. (42) is a better approximation to Eq. (41), we compare the effective potentials $V_{eff}$ versus $\frac{1}{r^2}$, $V_{eff1}$ versus approximation in Eq. (41) and $V_{eff2}$ versus the approximation in Eq. (42) for different values of $\alpha$ in Figs 4 – 6.

Substituting Eq. (42) into Eq. (40) and using the transformation

$$s = e^{-2\alpha r} \qquad (43)$$

Eq. (42) becomes

$$\frac{d^2U}{ds^2} + \frac{1}{s}\frac{dU}{ds} + \frac{1}{s^2}\left[-\varepsilon^2 - \gamma\frac{(a-s)}{(1-s)} + \frac{\theta s}{(1-s)} - \frac{\phi s}{(1-s)^2}\right]U = 0 \qquad (44)$$

where

$$-\varepsilon^2 = \frac{\mu E}{2\hbar^2\alpha^2} \qquad (45)$$

$$\gamma = \frac{\mu V_0}{2\hbar^2\alpha^2 b} \qquad (46)$$

$$\theta = \frac{1}{4\alpha^2}\left[\frac{2\mu V_1}{\hbar^2} - \omega\left(\frac{(D-1)(D-3)}{4} + l(l+D-2)\right)\right] \qquad (47)$$

$$\phi = \frac{1}{4\alpha^2}\left[\frac{2\mu V_2}{\hbar^2} + \lambda\left(\frac{(D-1)(D-3)}{4} + l(l+D-2)\right)\right] \qquad (48)$$

Equation (44) can further be reduced to the form,

$$\frac{d^2U}{ds^2} + \frac{(1-s)}{s(1-s)}\frac{dU}{ds} + \frac{1}{s^2(1-s)^2}[-(\varepsilon^2 + A)s^2 + (B + 2\varepsilon^2)s - (\varepsilon^2 + C)]U = 0 \qquad (49)$$

where

$$A = \gamma + \theta \qquad (50)$$

$$B = (a+1)\gamma + \theta - \phi \qquad (51)$$

$$C = \gamma a \qquad (52)$$

Comparing Eq. (49) and Eq. (12) and making further calculations, we obtain

$$c_1 = c_2 = c_3 = 1, \xi_1 = \varepsilon^2 + A, \ \xi_2 = B + 2\varepsilon^2, \xi_3 = C + \varepsilon^2, c_4 = 0$$

$$c_5 = -\frac{1}{2}, c_6 = \varepsilon^2 + A + \frac{1}{4}, c_7 = -B - 2\varepsilon^2, c_8 = \varepsilon^2 + C$$

$$c_9 = \sqrt{\phi + \frac{1}{4}}, c_{10} = 1 + 2\sqrt{\varepsilon^2 + \gamma a}, \ c_{11} = -2\sqrt{\phi + \frac{1}{4}},$$

$$c_{12} = \sqrt{\varepsilon^2 + \gamma a}, \ c_{13} = \frac{1}{2}\left(1 - \sqrt{1 + 4\phi}\right)$$

Using Eq. (24) we obtain the energy equation as,

$$\varepsilon^2 = \left[\frac{(1-a)\gamma + \theta - \phi - \left(n^2 + \frac{(2n+1)}{2}[1+\sqrt{1+4\phi}]\right)}{n + [1+\sqrt{1+4\phi}]}\right]^2 \tag{53}$$

From Eq. (49) we obtain the energy eigenvalues as

$$E_{n,l} = \frac{-2\hbar^2 \alpha^2}{\mu}\left[\frac{(1-a)\gamma + \theta - \phi - (n^2 + (2n+1)\sigma)}{2(n+\sigma)}\right]^2 + \frac{aV_0}{b} \tag{54}$$

where $\sigma = \frac{1}{2}\left[1 + \sqrt{1 + 4\phi}\right]$ \hfill (55)

## 5. Discussion

In this subsection, we consider some special cases of the potential in consideration

(i) If we set $V_0 = V_2 = 0, a = 0 \ and \ b = 1$. The potential in (38) reduces to [40]

$$V(r) = \frac{-V_0 e^{-2\alpha r}}{1 - e^{-2\alpha r}} \tag{56}$$

which is the Hulthen potential. Furthermore, if we set $D = 3$, $\omega = 0$ and $\lambda = 4\alpha^2$ in the eigenvalue equation (54) we obtain

$$E_{n,l} = \frac{-2\hbar^2\alpha^2}{\mu}\left[\frac{\mu V_0}{4\hbar^2\alpha^2(n+l+1)} - \frac{(n+l+1)}{2}\right]^2 \tag{57}$$

This is the eigenvalue of the Hulthen potential. The result is consistent with that of equation (36) in ref [35] if we substitute $\alpha = \frac{\delta}{2}$.

**(ii)** If we set $V_0 = V_2 = 0, b = 1$ and $a = -1$, the potential in Eq. (38) reduced to Rosen Morse potential [43]

$$V(r) = \frac{-V_0(1+e^{-2\alpha r})}{(1-e^{-2\alpha r})} = \coth(\alpha r) \tag{58}$$

Again, if we set $D = 3$, $\omega = 0$ and $\lambda = 4\alpha^2$, in Eq.(54), we obtain the eigenvalues of the Rosen-Morse potential as

$$E_{n,l} = \frac{-2\hbar^2\alpha^2}{\mu}\left[\frac{\mu V_0}{2\hbar^2\alpha^2(n+l+1)} - \frac{(n+l+1)}{2}\right]^2 - V_0 \tag{59}$$

Now using Eq. (29) we obtain the first part of the wave function as

$$\varphi(s) = s^\mu(1+s)^{\frac{1}{2}(-v+1)} \tag{60}$$

where $\mu = \sqrt{\varepsilon^2 + \gamma a}$ and $v = \sqrt{1 + 4\phi}$ and from Eq. (25) the weight function is

$$\rho(s) = s^{1+2\mu}(1+s)^{-v} \tag{61}$$

and following Eq. (26) the second part of the wave function is

$$\chi_n(s) = P_n^{1+2\mu,2v}(1+2s) \tag{62}$$

The Jacobi polynomials we documented in literatures [37–39]. The radial wave function is thus

$$U(r) = N_{n,l}(e^{-2\alpha r})^{\mu}(1 - e^{-2\alpha r})^{(v+1)} P_n^{1+2\mu,-v}(1 - 2e^{-2\alpha r}) \quad , \quad (63)$$

where $N_{n,l}$ is the normalization constant.

## 6. Numerical Results

The numerical results of the approximation schemes we employed are displayed in Fig.( 4-6). It shows that this approximation is good for all values of the parameter $\alpha$ since one can always adjust the parameters $\omega$ $and$ $\lambda$ unlike the approximation in equation (41) which is only good for small values of $\alpha$. We make appropriate choices for the parameters $\alpha, \omega, \lambda, V_0, V_1, b$ $and$ $a$ and computed the energy eigenvalues for the system for different values of $\omega$ $and$ $\lambda$ as shown in table (1-3) for different $n$ $and$ $l$.

## 7. Conclusions

In this paper, we investigate the solutions of the Schrödinger equation for the Eckart plus Hylleraas potentials in D-space, using a new approximation method for the centrifugal term via parametric Nikiforv-Uvarov method. The eigenvalues of the potential reduces to that of well known potentials viz Hulthen potential in Eq. (57) and Rose-Morse potential in Eq. (59), when we make appropriate choices of the parameters $\alpha, \omega, \lambda, V_0, V_1, b$ $and$ $a$.

Finally, we also obtain the wave function and expressed it in terms of the Jacobi polynomials.

**References**

1. O. Gomul and M. Kocak, Quant-Ph/0106144(2001)

2. K. J. Oyewumi, F. O. Akinpelu and A. D. Agboola, *Int. J. Theor. Phys* **47**(2008) 1039

3. A. N. Ikot and L. E. Akpabio,*Appl.Phys.Res.***2**(2010)202.

4. A. N. Ikot, L. E. Akpabio and E. B. Umoren, *J. Sci. Res* **3**(1) (2011) 25

5. H. Hassanabadi, S. Zurrinkamar and H. Rahimov, *Commun. Theor. Phys.* **56** (2011) 423

6. R. L. Greene and C. Aldrich, *Phys.Rev.A* **14** (1976)2363

7. A.N. Ikot, *Afr.Phys.Rev.***6**(2011) 221

8. A. P. Zhang, W. C. Qiang, Y. W. Ling, *Chin.Phys.Lett.* **26**(10) (2006) 100302

9. G. F. Wei, C. Y. Long, X. Y. Duan and S. H. Dong,*Phys.Scr.* **77**(2008) 035001

10. Y. P. Varshni, *Rev.Mod.Phys.***29** (4) (1957) 664

11. E. A. Hylleraas, *J.Chem.Phys.* **3**(1935) 595

12. J. L. Cardose and R. Alvarez – Nodarse, *J.Phys.A36* (2003) 2055

13. S. H. Dong and G. H. Sun, *Phys.Lett. A.***314** (2003) 261

14. S. H. Dong, C. Y. Chen and M. L. Cassou, *J.Phys.B* **38**(2005) 2211

15. K. J. Oyewumi and A. W. Ogunsola, *J.Pure Appl.Sci.***10**(2) (2004) 343

16. G. Paz, *Eur.J.Phys.***22**(2011)337

17. S. H. Dong, G.H. Sun and Popov, *J. Math. Phys.* **44**(10) (2003) 4467

18. S. Ikhadair and R. Sever, *Mol. Struc:THEOCHEM* **855** (2008) 13

19. S. H. Dong and Z. Q. Ma, *Phys. Rev.A* **65** (2002) 042717

20. A. N. Ikot, L. E. Akpabio and J. A. Obu, *Jour.Vect.Relat.***6**(2011) 1

21. A. N. Ikot, O. A. Awoga, L. E. Akpabio and A. D. Antia, *Elixir vib spec.* **33**(2011) 2303

22. A.N. Ikot, A. D. Antia, L. E. Akpabio and J. A. Obu, *Jour.Vect .Relat.* **6**(2) (2011)65

23. F. Yasuk, A. Durwins and I. Boztsun, *J.Math.Phys.* **47**(2006) 083202

24. B. J. Falaye and K. J. Oyewumi, *Afri.Rev.Phys.***6**(2011)0025

25. S.M. Ikhadair and R. Sever, *J.Math.Chem.***42** (3) (2007) 461

26. S. M. Ikhadair and R. Sever, *Int.J.Theor.* **46**(6) (2007) 1643

27. S. M. Ikhadair, *Chin.J.Phys.***46**(3) (2008) 291

28. S. M. Ikhadair and R. Sever, *Cent. Eur.J.Phys.* **6**(2008) 685

29. A. F. Nikiforov-Uvarov, *Special Functions of Mathematical Physics* (Birkhauser, Basel, 1998)

30. C. Tezcan and R. Sever, *Int .J.Theor.Phys.* **48**(2009) 337

31. A. Arda, R. Sever and C. Tezcan, *Chin.J.Phys.* **48**(1) (2010) 27

32. H. Hassanabadi, S. Zurrinkamar and A. A. Rajabi, *Commun.Theor. Phys.* **55**(2011) 541

33. S. H. Dong, W. C. Qiang, G. H. Sun and V. B. Bezerra, *J.Phys.A: Math.Theor.* **40**(2007)10535

34. O. Bayrak, G. Kocak and I. Boztosun, *J.Phys A: Math.Gen.* **39**(2006) 11521

35. F. Taskin and G. Kocak, *Chin.Phys.B* **19**(9) (2010) 090314

36. S. M. Ikhadair, *Eur.Phys.Jour. A.* **39**(2009) 307

37. M. Abramowitz and I. A. Stegun, *Handbook of Mathematical Functions with Formulas, Graphs and Mathematical Tables,* (Washington, 1972)

38. A. D. Polyanin and A. V. Manzhirov, *Handbook of Integral Equations,* (Washington, 1998)

39. I. S. Gradshteyn and I. M. Rhyzhik, *Table of Integrals, Series and Product, (*Elsevier, Burlington, 2007).

40. W.C.Qiang and S.H.Dong,*Phys.Scr.***79(**2009)045004.

41. S.H.Dong and J.Garcia-Ravelo,*Phys.Scr.***75(**2007)307.

42. C.S.Jia,P.Guo and X.L.Peng,*J.Phys.A:Math.Gen.***39**(2006)7737.

43. S.Debnath and B.Biswas,*EJTP,***26**(2012)191.

Table 1: Energy eigenvalues of the potential for $\omega = 1.6 \text{ and } \lambda = 3.2$

| n | l | Eigenvalues | | |
|---|---|---|---|---|
| | | D=3 | D=4 | D=5 |
| 0 | 0 | -0.693561969 | -1.483794256 | -2.749854638 |
| 1 | 0 | -2.400662738 | -3.520372978 | -5.150960406 |
| 2 | 0 | -5.107768168 | -6.565715334 | -8.583619961 |
| | 1 | -8.583619961 | -11.07635272 | -14.02084003 |
| 3 | 0 | -8.81487444 | -10.61307885 | -13.02521012 |
| | 1 | -13.02521012 | -15.92716134 | -19.28351018 |
| | 2 | -19.28351018 | -23.08296764 | -27.32202956 |
| 4 | 0 | -13.52198098 | -15.66115049 | -18.47027444 |
| | 1 | -18.47027444 | -21.78663214 | -25.56228156 |
| | 2 | -25.56228156 | -29.78098662 | -34.43704278 |
| | 3 | -34.43704278 | -39.52876901 | -45.05614563 |
| 5 | 0 | -19.22908762 | -21.7095332 | -24.91696686 |
| | 1 | -24.91696686 | -28.65040322 | -32.84945913 |
| | 2 | -32.84945913 | -37.4928681 | -42.57256298 |
| | 3 | -42.57256298 | -48.08566295 | -54.0315293 |
| | 4 | -54.0315293 | -60.41053758 | -67.22351809 |

Here $\alpha = 1, \omega = 1.6, \lambda = 3.2, a = 2.0, b = 50, \mu = 1,$

$V_0 = 1.0 MeV, V_1 = 0.01 MeV, V_2 = 0.5 MeV$

Table 2: Energy eigenvalues of the potential for $\omega = 1.7$ and $\lambda = 3.3$

| n | l | Eigenvalues | | |
|---|---|---|---|---|
| | | D=3 | D=4 | D=5 |
| 0 | 0 | -113.1097402 | -560.9727952 | -1316.065556 |
| 1 | 0 | -201.8384501 | -694.656059 | -1491.095494 |
| 2 | 0 | -315.5719709 | -856.6308917 | -1697.976766 |
| | 1 | -1697.976766 | -2836.531982 | -4271.56472 |
| 3 | 0 | -454.3074065 | -1045.392669 | -1934.208287 |
| | 1 | -1934.208287 | -3118.419739 | -4597.997338 |
| | 2 | -4597.997338 | -6373.079767 | -8443.782235 |
| 4 | 0 | -618.0437527 | -1260.207686 | -2198.334215 |
| | 1 | -2198.334215 | -3429.966834 | -4955.615228 |
| | 2 | -4955.615228 | -6775.822056 | -8890.971435 |
| | 3 | -8890.971435 | -11301.32424 | -14007.05962 |
| 5 | 0 | -806.7805866 | -1500.683512 | -2489.460427 |
| | 1 | -2489.460427 | -3769.946194 | -5342.999737 |
| | 2 | -5342.999737 | -7209.520922 | -9370.16336 |
| | 3 | -9370.16336 | -11825.38215 | -14575.49657 |
| | 4 | -14575.49657 | -17620.73484 | -20961.26357 |

Here $\alpha = 5, \omega = 1.7, \lambda = 3.3, a = 2.0, b = 50, \mu = 1,$

$V_0 = 1.0 MeV, V_1 = 0.01 MeV, V_2 = 0.5 MeV$

Table 3: Energy eigenvalues of the potential for $\omega = 12$ and $\lambda = 3.1$

| n | l | Eigenvalues | | |
|---|---|---|---|---|
| | | D=3 | D=4 | D=5 |
| 0 | 0 | - | - | -0.006591882 |
| 1 | 0 | -0.051095051 | -0.056748905 | -0.066201555 |
| 2 | 0 | -0.151319951 | -0.1577294 | -0.168286976 |
| | 1 | -0.168286976 | -0.18286666 | -0.201369665 |
| 3 | 0 | -0.291655888 | -0.298908764 | -0.310765078 |
| | 1 | -0.310765078 | -0.326970774 | -0.347291388 |
| | 2 | -0.347291388 | -0.371548105 | -0.399625631 |
| 4 | 0 | -0.472024074 | -0.480146418 | -0.493359927 |
| | 1 | -0.493359927 | -0.51130219 | -0.533626149 |
| | 2 | -0.533626149 | -0.560049273 | -0.590366077 |
| | 3 | -0.590366077 | -0.62444112 | -0.662195555 |
| 5 | 0 | -0.692404982 | -0.701407183 | -0.71600143 |
| | 1 | -0.71600143 | -0.735725353 | -0.760129567 |
| | 2 | -0.760129567 | -0.78883907 | -0.821569698 |
| | 3 | -0.821569698 | -0.858120539 | -0.89835829 |
| | 4 | -0.89835829 | -0.94220149 | -0.989607378 |

Here $\alpha = 5, \omega = 12, \lambda = 3.1, a = 2.0, b = 50, \mu = 1$,

$V_0 = 1.0 MeV, V_1 = 0.01 MeV, V_2 = 0.5 MeV$

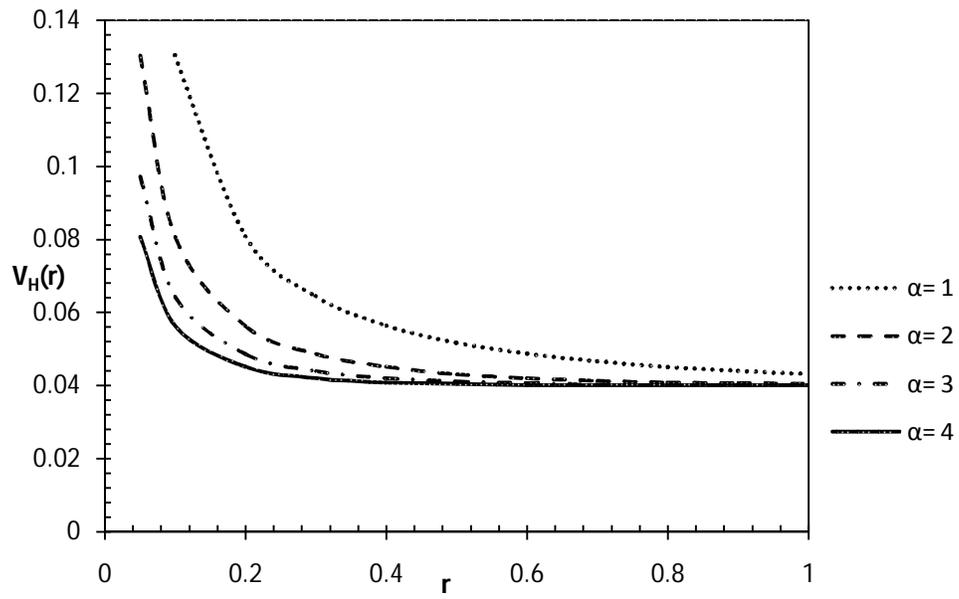

**Fig. 1: A plot of modified deformed Hylleraas potential as a function of r with a=2.0, b=50.0, V$_o$=1.0MeV for various values of the parameter α=1, 2, 3 and 4**

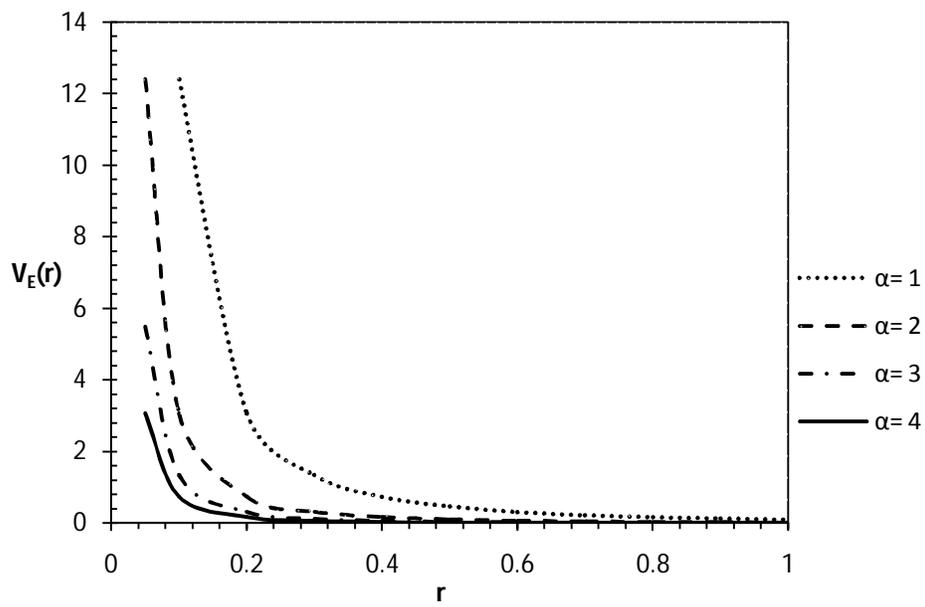

**Fig. 2:** A plot of Eckart potential as a function of r with $V_1$=0.01Mev, $V_2$=0.5Mev for various values of the parameter α=1, 2, 3 and 4

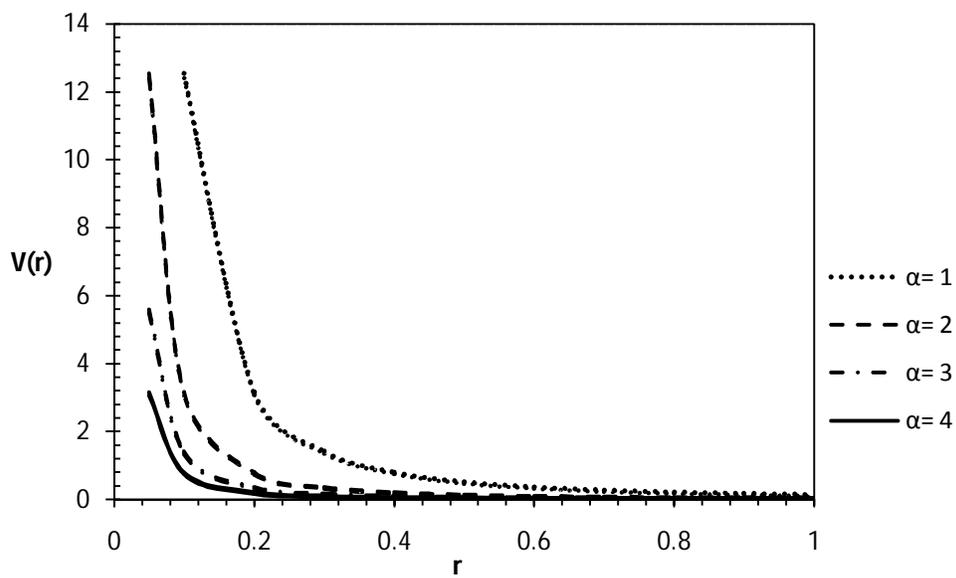

Fig. 3: A plot of the Deformed Hylleraas plus Eckart potentials as a function of r with $V_0$=1.0Mev, $V_1$=0.01Mev, $V_2$=0.5Mev, a=2.0, b=50.0 for various values of the parameter α=1, 2, 3 and 4

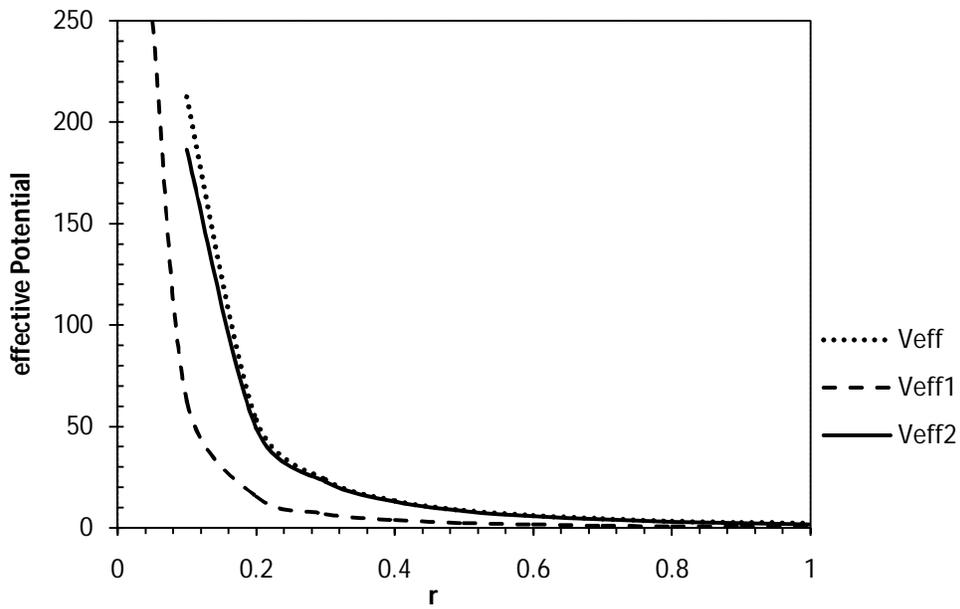

**Fig. 4:** A plot of the effective potentials ($V_{eff}$ for $1/r^2$, $V_{eff1}$ for the approximation in (41) and $V_{eff2}$ for the approximation in (42)) as functions of r with $V_0$=1.0Mev, $V_1$=0.01Mev, $V_2$=0.5Mev, a=2.0, b=50.0, l=1, $\alpha$=1, $\lambda$=3.2 and $\omega$=1.6

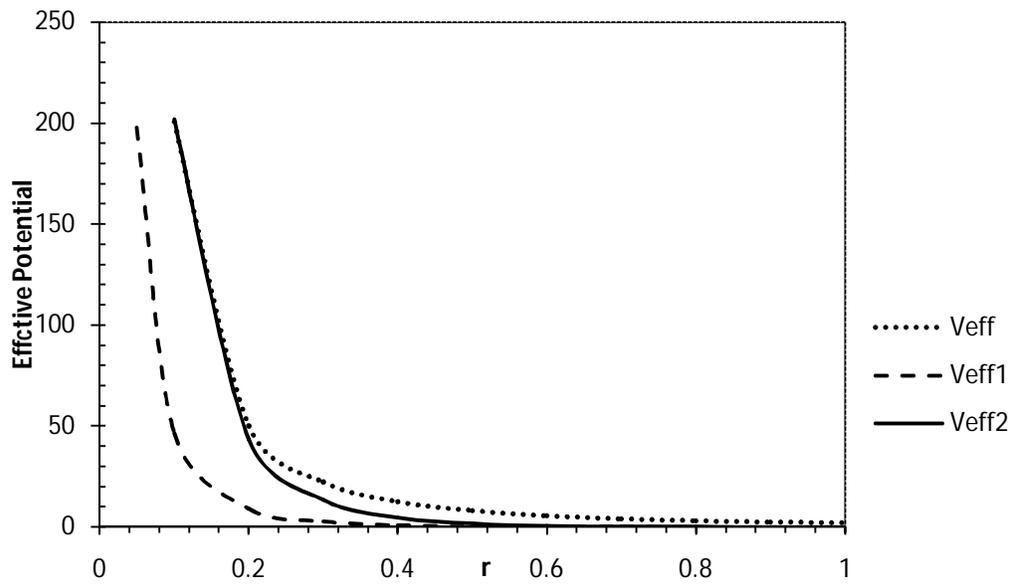

Fig. 5: A plot of the effective potentials ($V_{eff}$ for $1/r^2$, $V_{eff1}$ for the approximation in (41) and $V_{eff2}$ for the approximation in (42)) as functions of r with $V_0$=1.0Mev, $V_1$=0.01Mev, $V_2$=0.5Mev, a=2.0, b=50.0, l=1, $\alpha$=5, $\lambda$=3.3 and $\omega$=1.7

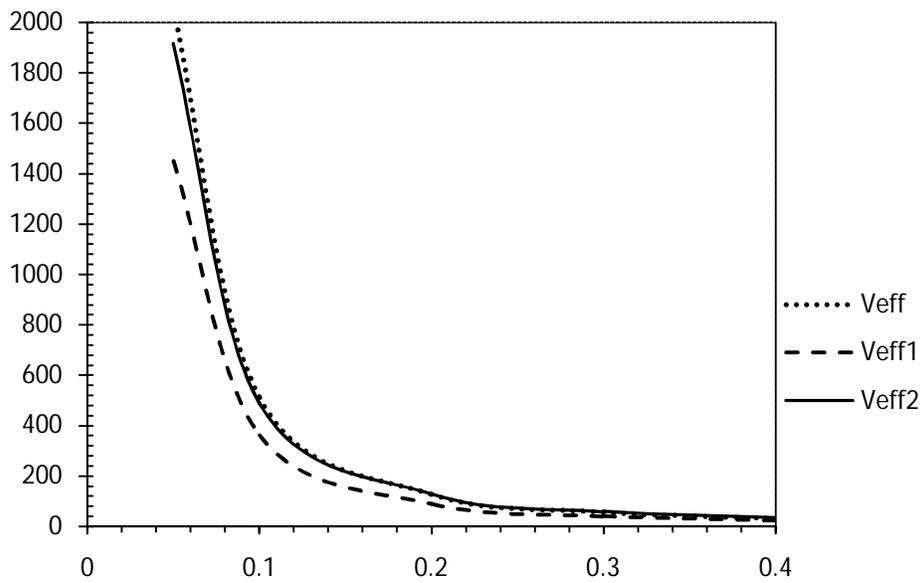

**Fig. 6: A plot of the effective potentials ($V_{eff}$ for $1/r^2$, $V_{eff1}$ for the approximation in (41) and $V_{eff2}$ for the approximation in (42)) as functions of r with $V_0$=1.0Mev, $V_1$=0.01Mev, $V_2$=0.5Mev, a=2.0, b=50.0, l=1, $\alpha$=0.2, $\lambda$=3.1 and $\omega$=12**